\documentclass[12pt]{iopart}
\usepackage{iopams}  
\usepackage{graphicx}
\usepackage[dvipsnames]{xcolor}

\begin{document}
	
	
	\title{Gravitational waves in the extended theory of gravity}
\author{Sourav Roy Chowdhury}

\address{Research Institute of Physics, Southern Federal University, 344090 Rostov on Don, Russia.}
\ead{roic@sfedu.ru}
	
\author{Maxim Khlopov}
\address{Research Institute of Physics, Southern Federal University, 344090 Rostov on Don, Russia.\\ National Research Nuclear University MEPhI, 115409 Moscow, Russia.\\Université de Paris,	CNRS, Astroparticule et Cosmologie, F-75013 Paris, France} 
\ead{khlopov@apc.in2p3.fr}

	\date{\today}
	
	\begin{abstract} 
		Extended Theories of Gravity are considered as a new approach for solving the infrared and ultraviolet scale problems the Standard Theory of Gravity (General Relativity). Observational evidence of gravitational waves and subsequent identification of the number of existing polarizations are an effective tool for testing general relativity and extended theories of gravity. The Newman-Penrose method is used to characterize the polarization modes for specific forms of $ f(R) $ in the present study. Both the forms of the $ f(R) $ theory belong to far more general variational class of gravitational waves, capable of presenting up to six separate polarizations states. We have introduced a new $ f(R) $ gravity models as an attempt to have a theory with more parametric regulations so that the model can be used to describe existing issues and discover different directions in gravity physics. The primary goal of this research is to look into the properties of gravitational waves with new cases. The model shows the existence of scalar degrees of freedom in $ f(R) $ gravity metric notation. 
		
	\end{abstract}


	\maketitle
	
	
	\section{Introduction}	

Modified Theories of Gravity are a new way that can be considered an extension of General Relativity (GR) to give an insight to the non resolved problems. They are a method for addressing theoretical and experimental problems in astrophysics, cosmology, and High Energy Physics by being consistent with the undeniably positive outcomes of Einstein's theory. The purpose is to include challenges like inflation, dark energy, dark matter, large scale structure, and, most importantly, to allow at least an impactful characterization of Quantum Gravity in a self-consistent arrangement \cite{Nojiri3,capozziello1,bondi}. 

Gravitational waves (GWs) have the potential to test any gravitational theory. Several alternative approaches to general relativity have been investigated so far. The main concerns of Einstein's general relativity are, 1) a compatible (i.e., unitary and renormalizable) theory of gravity does not exist yet. 2) In the quantization program for gravity, two approaches have been used: the covariant and the perturbative approaches. They do not point to a comprehensive theory of Quantum Gravity. 3) In the low-energy regime (concerning the Planck energy) at large scales, GR can be generalized by Hilbert–Einstein action terms of higher-order invariants and non-minimal couplings between matter and gravity \cite{capozziello,Nojiri,Nojiri2}. These lead, at the one-loop level, to a consistent and renormalizable theory. On the other hand, through suitable conformal transformations, mathematical equivalence between models with variable gravitational coupling and Einstein gravity have also been studied \cite{Dicke,Damour,Damour1}. Several researchers argued an actual physical distinction between the Jordan frame and the Einstein frame, citing experimental and observational evidence that the Jordan frame is better suited to matching solutions to data. The problem should be approached from a broader perspective, and the Palatini approach to gravity is widely discussed for modified theory, having several useful applications considered in \cite{Buchdahl,Olmo}. 

The remarkable feature of these Effective theories of gravity is that the field equations are fourth-order and, therefore, more involved than GR. Due to the higher-order terms, these field equations admit a much wider variety of solutions than the Einstein equations \cite{Ruzmaikina,Berkin}. The quadratic corrections to the Hilbert–Einstein theory provides exciting cosmology \cite{neto,Starobinsky}. Some of the modified views predict different propagation speeds and polarization modes of GWs from the idea of general relativity. Thus, the study of GWs is a new and overwhelming method to test any theory of gravity. Also, if we cannot find any deviations from the theoretical predictions of general relativity, we can obtain restrictions on the modifications. For example, the velocity of GW propagation provides us with limitations on several classes of modified gravities.

In this context, the $ f(R) $ theory of gravity is one of the well-appreciated modified gravity theories accepted by the scientific community over a significant time. x It is considered as a hierarchical theory over general relativity. Several viable $ f(R) $  gravity models are compatible with observational constraints coming from GWs and other astrophysical probes. According to Starobinsky \cite{Starobinsky}, the higher order terms may resemble a cosmological constant. 

This gravity model shows the scalar degrees of freedom in $ f(R) $  gravity metric formalism. The theory includes a scalar mode of polarization of GWs as a result of this. This polarization mode remains in a mixed state, with one being a massive longitudinal mode and the other being a transverse massless breathing mode with non-vanishing trace. The massive longitudinal mode travels slower than the standard tensor modes found in GR \cite{Corda,Corda1,Liang}. In general, in terms of Riemman tensor $ R_{tjtk} $ these modes can be represented. $ P_{\times}=R_{txty} $ represents cross mode where as $ P_+=R_{txtx}+R_{tyty} $ represents the plus mode, the vector-x mode is represented by $ P_{xz}=R_{txtz} $ and the vector-y mode by $ P_{yz}=R_{tytz} $, and the longitudinal mode is depicted by $ P_l=R_{tztz} $ \cite{hyun,Chen}. Furthermore, the potential and mass of scalar gravitons in both Jordan and Einstein frames to obtain a greater understanding of the extended gravity theory. Different models explain both primordial and current dark energy and meet the solar system tests along with imposed restrictions on the model \cite{Chen,Amendola,Amendola1,Bamba}. 

Alves et al. \cite{Alves2} analysed the $ f(R) $ framework using GWs polarisation. Along with other $ f(R) $ theoretical models, the model verifies the effectiveness of scalar degrees of freedom in $ f(R)  $ gravity metric methodology. In theory, there is a scalar mode of polarisation of GWs. This polarisation mode has two states: a massive longitudinal mode and a transverse massless breathing mode with non-vanishing trace \cite{gogoi}. For the higher order of extended gravity $ (f(R) = R + \alpha R^2) $, Capozziello and Laurentis \cite{capozziello} explore the Palatini representation, conformal transformations, and new polarisation states for gravitational radiation. Taking into consideration the chameleon mechanism, the scalar mode of GWs emerges in the framework of $ f(R) $ gravity by Katsuragawa et. al \cite{Katsuragawa}. To understand the gravitational field, diferent extended theories such as $f(R)$ gravity, scalar–tensor gravity, Gauss–Bonnet gravity, as well as other gauge theories are considered. The Teleparallel Equivalent General Relativity (TEGR) are  taken into account which emphasized that the GR may be formulated in terms torsion instead of curvature \cite{Bajardi}.

We have introduced a couple of $ f(R) $ gravity model in this paper to provide a potential theory with a number of controlling parameters so that it can be phenomenologically used to describe the existing observational issues and explore different directions in gravity physics. In Sec. (\ref{s1}), the basic structure and the field equations are constructed. On the basis of the functional forms of $f(R)$, Sec. (\ref{s2}) discusses the associated scalar field and properties. We calculated the Newman-Penrose (NP) quantities in Sec. (\ref{s3}) to characterize the polarization modes. Sec. (\ref{s4}) explained and concluded about GWs and the polarization modes corresponding to our model.

\section{Basic formalism}\label{s1}

In the framework of extended gravity the total modified action can be define as,

\begin{equation}
S = \frac{1}{2 k} \int d^4x \sqrt{-g} f(R) + \int d^4x \sqrt{-g} \mathcal{L}_m,
\end{equation}

where, $ f(R) $ is the function of Ricci scalar $ R $, and $ k $ is the coupling constant.

The action of the field, with $ -g $ being the determinant of the metric. Throughout this article, we use the natural units. 

The matter Lagrangian  is directly related with the stress-energy tensor as follow 

\begin{equation}
T_{\mu \nu} = -\frac{2}{\sqrt{-g}} \frac{\delta (\sqrt{-g} \mathcal{L}_m )}{\delta g^{\mu \nu}}.
\end{equation}

and its trace by $T = T_{\mu \nu}g^{\mu \nu}$, respectively. We restricted  that the Lagrangian density $ L $ is only dependent on the metric tensor components $g^{\mu \nu}$, and is not-dependent on its derivatives.          

On varying the action we find,
\begin{eqnarray}
\delta S = \frac{1}{2 k}  \int \Big[ \left( f'(R)  R_{\mu \nu} -\frac{1}{2}g_{\mu \nu}f(R) \right) & & \delta g^{\mu \nu} + f'(R) \delta R_{\mu \nu} g^{\mu \nu} \nonumber  \\ & &+ \frac{1}{\sqrt{-g}} { \frac{\delta( (\sqrt{-g} \mathcal{L_\phi} ))}{ \delta g^{\mu \nu}}} \Big] \sqrt{-g}d^4x,
\end{eqnarray}

From the variation Eq , the field equation for $ f(R) $ gravity as follow,
\begin{equation}
f'(R) R_{\mu \nu} - \frac{1}{2} f(R) g_{\mu \nu} - \nabla_\mu \nabla_\nu f'(R) + g_{\mu \nu} \Box f'(R) = kT_{\mu \nu}.
\end{equation}

The generalized form of the Ricci tensor in the framework of extended gravity can be written as follows
\begin{equation}
R_{\mu \nu} = \frac{1}{f'(R)} \left[ \frac{1}{2} f(R) g_{\mu \nu} + \nabla_\mu \nabla_\nu f'(R) - g_{\mu \nu} \Box f'(R) + kT_{\mu \nu} \right].\label{e5}
\end{equation}

And the corresponding Ricci scalar is
\begin{equation}
R = \frac{1}{f'(R)} \left[ 2 f(R)- 3~ \Box f'(R) + kT \right]. \label{e6}
\end{equation}

This gives us the dynamic equation for the Ricci scalar, though it doesn't follows the notion of GR, i.e. $ R = -kT $ additional modified terms are presented in the equation. 

Following \cite{capozziello}, we can define $ \psi = f'(R) $. 

Solving the above relation, one can express the Ricci scalar in terms of $ \psi $ and correlate with the potential.

$ f''(R) > 0 $ provides the stable cosmological perturbation and also the positivity on the scalar modes of GWs.  Again $ f'(R) > 0 $ leads to positive effective gravitational coupling.

\section{ Modified gravity : New ansatzs $ f(R) $} \label{s2}

The $ f(R) $ theoretical model with some parametric control seems to be more appropriate in fitting with the observational data by putting constrains on its parameters. In the metric representation, we presently look at the variational principle and the field equations of $ f(R) $ gravity. The field equations in these extended theories are hierarchically more general, thus spanning more possibilities than GR (which can be regained as the special case $ f(R) = R $). The field equations admit a much extended range of solutions than the Einstein equations due to increased order.

Throughout this paper, we worked with the waves propagating in vacuum. We set $ kT_{\mu \nu}  = 0$.

\subsection{Case-I}

We present a new $ f(R) $ gravity model with two additional parameters. 
\begin{equation}
f(R)  = R - \alpha \cos(\beta R), 
\end{equation} 

where $ \alpha $ and $ \beta $ are positive coupling constants. Second term can be considered as a perturbative correction. 

We take the form of $ f(R) $ such that the second derivative is positive. For this case, $ f''(R) = \alpha \beta^2 \cos(\beta R) $. The model shows stable cosmological perturbation for positive $ \alpha $.

Following Eq. \ref{e5}, the Ricci tensor for the model is 
\begin{eqnarray}
R_{\mu \nu} = \frac{1}{1+\alpha \beta \sin(\beta R)} \Big[ \frac{1}{2} R g_{\mu \nu} - \alpha \cos(\beta R)  + \alpha \beta & & \nabla_{\mu} \nabla_{\nu} \sin(\beta R)\nonumber\\ & & - g_{\mu \nu} \alpha \beta \Box \sin(\beta R) \Big] \label{e8}
\end{eqnarray}

The dynamical equation of Ricci scalar obtained by contracting Eq. \ref{e8}.
\begin{equation}
\Box \sin(\beta R)+\frac{2}{3 \beta} \cos(\beta R) - \frac{1}{3} R \sin(\beta R)- \frac{R}{3  \beta} = 0 \label{e9}
\end{equation}

With restriction that the system is bounded only for first order terms and our is $ R << 1$ (Minkowski), the solution of the above equation can be expected in the following form,
\begin{equation}
R = \frac{2 \alpha}{9} + R_o \exp(ik_\rho x^\rho),~~k_\rho k^\rho = \frac{3}{\alpha \beta^2},
\end{equation} 

where, $ R_o $ is a constant.

\begin{figure}[h!]\centering
	\includegraphics[width=8.9cm]{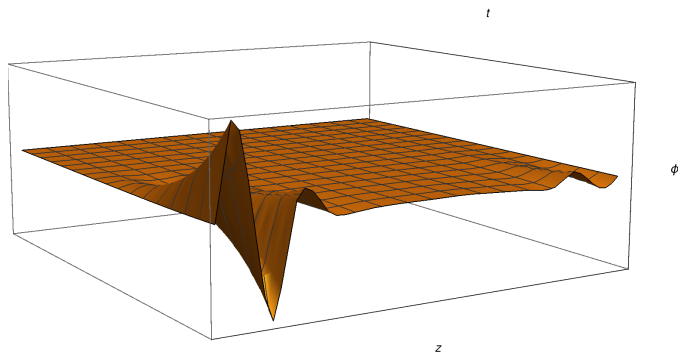}
	\caption{Propagation ($\phi$)  for the perturbation of vacuum scalar field. For the propagation we choose $\alpha = 1 $ and $ \beta = 0.5 $.}\label{f1}
\end{figure}

\subsection{Case-II}

We are also interested in the phenomenology for the following form
\begin{equation}
f(R) = R - R_c \beta\left[1-\left(1+\frac{R}{R_c}\right)^{-n} \right], \label{e10}
\end{equation} 

where  $ \beta $ is a positive coupling constant, and $ R_c $ is the characteristic curvature constant, of the same dimension as of Ricci scalar.

Two variables and $ R_c $ are used to calculate the correction factor. It is based on the non-linear $ f(R) $ gravity model.
We will show that this model meets the basic criteria for a viable $ f(R) $ gravity model.

Substituting the form the gravity in Eq. \ref{e5} we obtain the ollowin relations between $ R_{\mu \nu} $ and the Ricci scalar $ (R) $,
\begin{eqnarray}
R_{\mu \nu} = & &\frac{1}{1- \beta n \chi ^{-(n+1)}} \Bigg[ \frac{1}{2} R g_{\mu \nu} - \frac{1}{2} \beta R_c g_{\mu \nu} \left\{ 1- \chi ^{-n} \right\}  + \beta n \nabla_{\mu} \nabla_{\nu} \chi ^{-(n+1)} \nonumber \\ & & \hspace{7cm} - g_{\mu \nu}  \beta n \Box \chi ^{-(n+1)} \Bigg] \label{e11}
\end{eqnarray}

We have a dynamical equation for $ R $, which can be obtained directly on contracting Eq. \ref{e11}, as follow
\begin{eqnarray}
R = & &\frac{1}{1- \beta n \chi ^{-(n+1)}} \Bigg[ 2 R - 2 \beta R_c \left\{ 1- \chi ^{-n} \right\}  - 3 \beta n \Box \chi ^{-(n+1)} \Bigg] \label{e12}
\end{eqnarray}

We have considered, $ \chi = \left( 1+ \frac{R}{R_c}\right) $

with the solution,
\begin{equation}
R = R_o \exp(ik_\eta x^\eta),~~ k_\eta k^\eta=\frac{(9-5\beta n)R_c}{3\beta n(n+1)}
\end{equation}
where, $ R_o $ is a constant.
For our model 

\begin{figure}[h!]\centering
	\includegraphics[width=8.9cm]{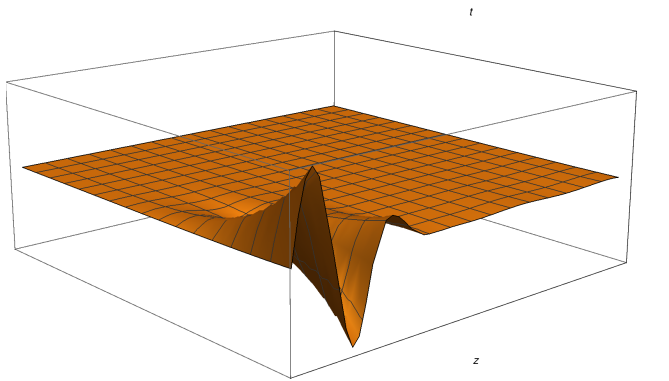}
	\caption{Propagation ($\phi$) for the perturbation of vacuum scalar field. For the propagation we choose $R_c = 1 $ and $ \beta = 0.2 $.} \label{f2}
\end{figure}

\begin{equation}
f''(R) = \beta n (n+1) \frac{1}{R_c}\left(1+\frac{R}{R_c}\right)^{-(n+2)} > 0
\end{equation}

for both $ \beta $ and $ n $ are  positive. 

It follows the stable cosmological perturbation.

\section{Polarization modes}\label{s3}

Newman-Penrose (NP) \cite{New1, New2} formalism gives us the extra polarization modes;  The references  \cite{eardley1, eardley2} highlight the matter with more intricate details.
NP quantities can be defined out of the tetrads $(e_t, e_x, e_y, e_z)$ for the system at each point, agreeing to the six distinct polarization modes of Gravitational waves. Such vectors can be portrayed as the NP tetrads as $k,~l,~m~,\bar{m}$. The real and imaginary null vectors are, 
\begin{eqnarray}
k = \frac{1}{\sqrt{2}}(e_t+e_z),~~
l = \frac{1}{\sqrt{2}}(e_t-e_z),\\
m = \frac{1}{\sqrt{2}}(e_x+ie_y),~~
\bar{m}= \frac{1}{\sqrt{2}}(e_x-ie_y).\\
-k.l=m. \bar{m}=1, ~~E_a = (k, l, m, \bar{m}).\nonumber
\end{eqnarray}

The other possible inner products become zero.

The independent parts of the Riemann tensor $R_{\lambda \mu \kappa \nu}$ are rendered by nine components of the traceless Ricci tensor ($\Phi$'s), ten components of the Wely tensor ($\Psi$'s), and a curvature scalar ($\Lambda$) in the NP formalism. They are reduced to six components admitting symmetry properties. They are designated by $\Psi_2, \Psi_3,\Psi_4, \Phi_{22}$ real and  $ \Psi_3,\Psi_4 $ complex. The components of the Riemann tensor in the null tetrad basis are expressed in terms of these NP quantities in the following way,
\begin{eqnarray}
& &\Psi_2 = -\frac{1}{6} R_{lklk} \sim  \textrm{longitudinal scalar mode,}\nonumber\\
& &\Psi_3 = -\frac{1}{2} R_{lkl \bar{m}} \sim  \textrm{vector-x \& vector-y modes,}\nonumber\\
& &\Phi_{22} = - R_{lml\bar{m}} \sim  \textrm{breathing scalar mode.} \label{e30}\\
& &\Psi_4 = - R_{l\bar{m}l\bar{m}} \sim  \textrm{+,} \times \textrm{tensorial mode,}\nonumber\label{e32}
\end{eqnarray}

The other non zero NP variables are defined in terms of them as $\Phi_{11} = 3 \Psi_2 / 2$, $\Phi_{12}=\Phi_{21}= \Psi_3$ and $\Lambda = \Psi_2/2$, respectively. 

The Four NP variables  $\Psi_2, \Psi_3,\Psi_4, \Phi_{22}$  are classified based on their transformation properties in the Lorentz group for massless particles, the group E(2).  According to the transformation properties, the amplitudes of the four NP variables are not independent of the observer concerned. However $ \Psi_2 $ is invariant. The NP variables that do not contribute (have zero amplitude) continue to be zero independent of the observer.

The Reimann tensor and the Ricci scalars are related by:
\begin{eqnarray}
& & R_{lklk}= R_{lk},\nonumber \\
& & R_{lklm}= R_{lm},\nonumber\\
& & R_{lkl \bar{m}} = R_{l \bar{m}},\nonumber\\
& & R_{l \bar{m} l \bar{m}}= \frac{1}{2} R_{ll},\nonumber\\
& & R = -2R_{lklk}= 2R_{lk}.\label{e31}
\end{eqnarray}

\subsection{Case-I}

Considering the solution of Eq. \ref{e11}, with the constrains $ R<<1 $, we find the non-null components as follow
\begin{eqnarray}
& & R_{tt} = \frac{1}{(1+\alpha \beta^2 R)} \left[ - \frac{\alpha}{18} +  \frac{R_o}{2} \exp(ik_\rho x^\rho) \left\{ 1-2\alpha \beta^2 \left( k^2 - \frac{ 3}{\alpha \beta^2} \right) \right\} \right] \nonumber \\
& & R_{zz} = \frac{1}{(1+\alpha \beta^2 R)} \left[ \frac{\alpha}{18} -  \frac{  (1+2\alpha \beta^2 k^2) }{2} R_o \exp(ik_\rho x^\rho) \right]\nonumber \\
& & R_{tz} = \frac{1}{(1+\alpha \beta^2 R)} \left[  \alpha \beta^2 k \left( k^2 - \frac{ 3}{\alpha \beta^2} \right)^{1/2} R_o \exp(ik_\rho x^\rho) \right]
\end{eqnarray} 

From the above non-null quantities, we can state that the non zero NP quantities are

$ \Psi_2 \neq 0 $; $ \Psi_3 = 0 $ ; $ \Psi_4 \neq 0 $ and $ \Phi_{22} \neq 0 $\\

Therefore the existing polarizations are +, $ \times $ tensorial mode, breathing scalar mode and longitudinal scalar mode.

\subsection{Case-II}

For the second type of $ f(R) $ model, the non-null components are
\begin{eqnarray}
& & R_{tt} = \frac{R}{1- \beta n \chi ^{-(n+1)}} \left[ \frac{1}{2} - \frac{\beta n }{6} - \frac{\beta n (n+1)}{R_c} \left\{ k^2 - \frac{(9-5\beta n)R_c}{3\beta n(n+1)} \right\} \right] \nonumber\\
& & R_{tz} = -\frac{R}{1- \beta n \chi ^{-(n+1)}} \left[  \frac{\beta n (n+1)}{R_c} \left\{ k^2 - \frac{(9-5\beta n)R_c}{3\beta n(n+1)} \right\}^{1/2}k \right] \nonumber\\
& &R_{tt} = \frac{R}{1- \beta n \chi ^{-(n+1)}} \left[ -\frac{1}{2} + \frac{\beta n }{6} - \frac{\beta n (n+1)}{R_c} k^2  \right] \nonumber\\
\end{eqnarray}

As a result, using the method outlined in Eq. \ref{e32} and Eq. \ref{e31}, we can conclude the non null NP quantities are

$ \Psi_2 \neq 0 $; $ \Psi_3 = 0 $ ; $ \Psi_4 \neq 0 $ and $ \Phi_{22} \neq 0 $\\

Thus we get four polarization modes for the GW: +, $ \times $ tensorial mode, breathing scalar mode and longitudinal scalar mode.

\section{Conclusion}\label{s4}

In this article, we introduce two new types of $ f(R) $ models with perturbative corrections, which are parametric regulative quantities. The stability of a scalar field is defined by $ f''(R)>0 $.

The longitudinal scalar mode and the perpendicular scalar mode (breathing mode) are represented by amplitudes of $ \Psi_2 $ and $ \Phi_{22} $, respectively. Since $ \Psi_2 $ is non-null, it is worth to noting that the E(2) classification is II$_6 $. Therefore, a Lorentz observer can always measure all existing modes among six polarization states. As a result, it is always possible to find a Lorentz observer who can measure all six polarisation states for this class. On the other hand, in the Palatini approach, the GWs only have two polarization states as in GR. 

Besides that, the approach we employ is fundamental and very reliable, allowing us some crucial information about the theories under consideration. An essential observational GW amplitude is $ \Psi_2 $. The $ f(R) $ models in the metric formalism, such as the one discussed here, would be endorsed if amplitude $ \Psi_2 $ could be identified. 

For first type of model if we consider the coupling constant $ \alpha  = 0$, the relation reduces to the general relativity theory.  From Eq \ref{e8} and  Eq. \ref{e9}, we obtain $ R_{\mu \nu} = 0 $ and $ R = 0 $. Thus, all the components of $ R_{tjtk} $  become zero, and the only existing polarization modes is +, $ \times $ tensorial mode. The same is also true for second type of model, for zero coupling constant ($ \beta $).  The E(2) classification for GR is N$_2$ .

We restricted ourselves to $ R<<1 $ to find the propagation variation and components of $ R_{tjtk} $ for the first type of model. The variation is shown in Fig. \ref{f1}. The second type of model is constraints on $ n $, the value of $ n \neq -1 $. There is no oscillatory behaviour or propagation for such value. The considered value of $ R_c =1 $ for though out the paper. The propagation is shown in Fig. \ref{f2}. From the nature of propagation, it is clear that the maximum perturbation is near the source point for both cases.

The $ ``k^2 = 0"$ mode corresponds to a massless spin two field with two independent polarizations and a scalar mode, whereas a massive spin two ghost mode  corresponds to a massive spin two field with five independent polarisation tensors and a scalar mode. We could set up an effective approach to determine the theory of gravity if we can identify the polarization states \cite{Corda2}. However, the polarization of GWs in some $ f(R) $ gravity models in the Palatini approach are almost the same as general relativity, but not in general.

This current model appears to be useful in confronting a range of open problems in astrophysics and cosmology. 

\section*{Acknowledgment}
The southern federal university supported the work of SRC (SFedU) (grant no. P-VnGr/21-05-IF). SRC is also thankful to Ranjini Mondol of IISc, Bangalore, for the fruitful discussion to improve the manuscript. 

\newpage

\section*{References}


\end{document}